# Cancer and electromagnetic radiation therapy: *Quo Vadis*?

Mersini Makropoulou
Physics Department, School of Applied Mathematical and Physical Sciences,
National Technical University of Athens, Zografou Campus, Athens, Greece.
E-mail: mmakro@central.ntua.gr

**Abstract**

In oncology, treating cancer with a beam of photons is a well established therapeutic technique, developed over 100 years, and today over 50% of cancer patients will undergo traditional X-ray radiotherapy. However, ionizing radiation therapy is not the only option, as the high-energy photons delivering their cell-killing radiation energy into cancerous tumor can lead to significant damage to healthy tissues surrounding the tumor, located throughout the beam's path. Therefore, in nowadays, advances in ionizing radiation therapy are competitive to non-ionizing ones, as for example the laser light based therapy, resulting in a synergism that has revolutionized medicine. The use of non-invasive or minimally invasive (e.g. through flexible endoscopes) therapeutic procedures in the management of patients represents a very interesting treatment option. Moreover, as the major breakthrough in cancer management is the individualized patient treatment, new biophotonic techniques, e.g. photo-activated drug carriers, help the improvement of treatment efficacy and/or normal tissue toxicity. Additionally, recent studies support that laser technology progresses could revolutionize cancer proton therapy, by reducing the cost of the needed installations.

The aim of this review is to present some laser-based future objectives for cancer radiation therapy, aiming to address the relevant advances in the ionizing and non-ionizing radiation therapy, i.e. protons and heavy ions therapy, as well as photodynamic targeted and molecular therapies.

## 1. INTRODUCTION

Today, one of the major health problems for mankind is cancer. Cancer is the second most common cause of death in the US, exceeded only by heart disease, accounting for nearly 1 of every 4 deaths [1]. In developed countries every year some 40,000 per 10 million inhabitants are diagnosed as having cancer, while the World Health Organization has projected cancer-related mortalities to rise to over 11 million by 2030 [2]. Obviously, the universal demand in oncology is early cancer diagnosis on one part, and to remove cancer or precancerous growths or to relieve symptoms of cancer in the case of disease. The armamentarium of cancer treatment includes surgery, radiotherapy and chemotherapy. Among them, treating cancer with a high-energy photons beam is a well established therapeutic technique, developed over 100 years, and today is the most widely used modern cancer treatment - over 50% of cancer patients in developed countries will undergo traditional X-ray radiotherapy to cure their disease. Typically, the higher the dose, the more effective the radiotherapy is in destroying cancer. However, ionizing radiation therapy has also side effects, as the high-energy photons releasing their energy into malignant cells can also lead to significant damage to healthy tissues surrounding the tumor. Therefore, in nowadays, advances in cancer radiation therapy stare at a shift in the electromagnetic radiation spectrum to non-ionizing photons. For example, the laser based surgery and new biophotonic techniques, such as photodynamic therapy and targeted nanomedicine with photo-activated drug carriers, are used to improve cancer treatment efficacy concomitantly reducing normal tissue toxicity.

The year 2015 it was proclaimed by the UN General Assembly and it was celebrated worldwide as the "International Year of Light", just four years after the 50th anniversary of the discovery of the laser radiation. During the first 50 years laser light period, both research developments and applications of laser radiation are impressive, particularly in medicine (in all areas: diagnosis, therapy, rehabilitation and biomedical research and technology). In oncology, there are as well several biophotonic applications and ongoing research efforts in both early diagnosis and imaging, as well as in novel treatment options. Based on the relevant realizations, several groups worldwide protract the idea that, after the predominance of ionizing radiation therapy in the $20^{th}$ century, the $21^{st}$ century raised as the Biophotonics era for life sciences. Is this an optimistic target for human health improvement? In an attempt to answer this and the rhetoric question set in the title of this work "Cancer and electromagnetic radiation therapy*: Quo Vadis*??", it is useful to review first the basic concepts of what is cancer and what is expected in our days from cancer radiation therapy.

## *What is cancer?*

Cancer is a group of diseases caused by normal cells changing so that they grow in an uncontrolled way. Cancer is characterized by this uncontrolled growth or cell proliferation, which causes a lump or a collection of cells called a tumor to form, and by the spread of abnormal cells in distant tissues (*metastasis*). If the tumor is not treated and the metastasis is not controlled, cancer can cause problems and may eventually lead to the patient's death. The origin of the word cancer is credited to the Greek physician Hippocrates (460-370 BC), who used the terms *carcinos* and carcinoma to describe non-ulcer forming and ulcer-forming tumors. Carcinos was a giant crab in Greek mythology and, later, the Roman physician Celsus (28-50 BC) translated the Greek term into *cancer*, the Latin word for crab, [3]. Furthermore, the word tumor is from the Latin verb *tumere*, meaning to bulge or to swell [4]. However, not all swellings are tumors (e.g., an inflammation or other lumps). As already mentioned, tumor originates from the transformation of normal cells into cancer cells (carcinogenesis, from the Greek words *carcinos*=crab and *genesis*=birth), which acquire the ability to become unresponsive to normal physiological controls. According to Tracqui [4], this simple statement embraces the extremely diverse and complex reciprocal dialogues that cells and molecules comprising a tumor engage through highly regulated interactions favoring malignancy. Definitely, carcinogenesis is a multistep process, involving genetic alterations, as for example mutations in the genes regulating cellular growth, inhibition of programmed cell death (apoptosis), activation of oncogenes and deactivation of tumor suppressor genes.

As cancer is not just one disease but many diseases, we discriminate more than 100 different types of cancer. Most cancers are named from the organ or type of cell in which they start - for example, cancer that begins in the colon is called colon cancer; cancer that begins in melanocytes of the skin is called melanoma. Furthermore, cancer types can be grouped into broader categories. Obviously, different types of cancer can behave very differently. Apart of the organ of tumor's origin, more generally, tumors are classified according to the embryonic origin of the tissue from which they derive. We know from the science of embryology that, during the third week of the embryonic stage, three germ cell layers develop from which all organs and tissues originate. The three embryonic germ layers are the endoderm, mesoderm, and ectoderm. The term carcinoma is used to denote cancers of endodermal (e.g., gut epithelia cancers) or ectodermal (e.g. from tissue of the body's surface, skin and mucosa) origin. Ectodermal tumors also include cancers occurring in organs derived from epithelial structures such as the gastro-intestinal tract, liver, pancreas, respiratory, and urogenital tract. Cancers of mesodermal origin (e.g., bones, muscle, blood cells, or other connective or supportive tissue) are called sarcomas. Leukemia (blood cell cancers) is also classed as mesodermal cancers. Carcinomas make up more than 90% of malignant tumors. For definitions of other cancer-related terms, the interested reader can see several European, e.g. [5], or American governmental websites, e.g. [6].

As we mentioned, the universal effort in oncology is to remove cancer or precancerous growths or to assure a palliative intervention, to relieve the symptoms of cancer. However, even after many years of clinical research and practice, there are several obstacles in the effective treatment of the cancer worldwide and this disease is still uncontrollable. We still need to elucidate several aspects of tumor biology, tumor growth and its impact in cancer treatment, tumor regression etc. Definitely, the fight of cancer demands a multidisciplinary effort to cover the complications of the disease.

## 2. Tumor growth and the impact in cancer treatment

Tumor growth is a very complex phenomenon. Additionally to carcinogenesis, tumor growth occurs because of the proliferation of the tumor cells and the development of supporting stroma and vasculature (by angiogenesis). More specifically, the cumulative effect of carcinogenesis and mutagenesis is malignant transformation of cells, rapid growth and proliferation of these malignant cells, amplified angiogenesis, invasion into organs and body cavities, and subsequent metastasis. As we mentioned briefly before, metastasis refers to the spread of the cancerous cells from the point of origin to other body parts and systems through the vascular and lymphatic circulation systems. Although some tumors (e.g. leukemias, ascites tumors) grow as cell suspensions, most tumors grow as solid masses of tissue. Solid tumors make up more than 90% of all of human cancers [7]. Solid malignant tumors are basically composed of neoplastic cells, collectively called parenchyma, and non-tumorous elements, the stroma. The stroma and the neo-vasculature that arises provide tumor cells with oxygen and nutrients necessary to survive and grow [8]. The neo-vasculature also allows for the removal of toxic waste products associated with cellular metabolism [9] and is one of the two vehicles for metastatic spread, e.g. through the bloodstream (the other one is through the lymph system).

The study of the tumor development and growth has several distinct features; the biological approach for the natural history and the evolution of a specific tumor, the therapeutic procedures for tumor shrinkage and necrosis, and the mathematical approach for these inter-related processes. As many review papers and books are describing the basic biological models of tumor growth, in this work we will just mention briefly on the mathematical models of tumor growth, based on mathematical simulation and expressions of the dependence of tumor size on time, and their impact in cancer treatment. Nowadays, it is generally accepted that mathematical models are an efficient way to complement the results of preclinical research and clinical trials. Beside animal models and cell lines, which are often used for preclinical studies, there are computer models (*in silico* models) encompassing analytical (mathematical) or stochastic methodology, as well as physical and engineering concepts representing the biological world [10].

Since cell division is a binary process, it can be expected that tumor growth would be an exponential function with the volume increasing as a semi-logarithmic function of time. This implies a constant time for the tumor to double in volume (*volume doubling time*). Small tumors often express this form of growth function but, as they get larger, the growth rate of a tumor usually declines (longer doubling time) due to nutrient deprivation and other conditions [11, 12]. For the reader further interested to the biophysical models of tumor growth, a comprehensive review was reported by Tracqui some years ago [4].

The fact that cancer is characterized by uncontrolled cell proliferation has stimulated biomathematicians to construct different cell proliferation models (continuous, discrete, deterministic, stochastic) describing growth and kinetics of abnormal cell multiplication. Late 90', collaborating with the Greek "*In Silico Oncology Group*", led by Stamatakos, we reported our efforts to model *in vitro* tumor spheroids growth for small lung cell carcinoma and the impact of radiotherapy, with spatial visualization and a simplified simulation method [12, 13]. Certainly, over the last decades considerable efforts have been made worldwide in order to simulate tumor growth and the impact to various anti-cancer treatment schemes. The work on simulation included some of the most advanced mathematical medical science, such as stochastic cellular automata, discrete event simulation, hypermatrices and discrete operators [14]. A further step towards shaping an emerging analytical-computational discipline of ''*In Silico* Radiation Oncology'' performed by Stamatakos and his group with the elaboration of a four-dimensional, patient individualized*, in vivo* simulation model of glioblastoma multiforme tumor response to radiotherapy of [15]. Five years ago, the organizers of the "4th International Advanced Research Workshop on *In Silico* Oncology and Cancer Investigation", hold in Athens - Greece, described in the scope of the workshop that: "*Cancer is a natural phenomenon and as such it should be amenable to mathematical and computational description. Clinically driven complex multiscale cancer models can produce rather realistic spatio-temporal simulations of concrete clinical interventions such as radio-chemotherapy applied to individual patients. Clinical data processing procedures and computer technologies play an important role in this context* [16]". In this framework, a Euro-Japanese effort to develop the "Oncosimulator" was funded by the European Commission (EC) project "Advancing Clinico-Genomic Trials on Cancer" [14], while the clinical adaptation and validation of the multiscale modeling component of the system is in progress [17].

Coming back to the natural history and the biology of a tumor, we recall that the growth of a tumor is controlled by oxygen and nutrient supply (for example, nutrients as amino acids and glucose). Solid tumors begin as avascular polyps dependent upon the diffusion of oxygen and nutrients across the tumor surface. Further growth depends on the recruitment and proliferation of blood vessels through angiogenesis and is fueled by metabolic resources in the host environment [18]. In their comprehensive theory, Herman *et al* [18] presented a general quantitative framework that captures many of the essential features of tumor vascularization and growth, and how these are influenced by the host organism in an allometric theory to tumor growth modeling. There is abundant evidence that when a tumor has grown to a detectable size, its central region usually contains only dead cells and, therefore this central region of dead cells is called *necrotic core* of the tumor. In figure 1, a schematic representation of cross-section of a multi-cellular spheroid solid tumor is illustrated, based on the Greenspan mathematical model to mimic the growth of spherical necrotic tumors [19]. In this scheme, we can distinguish a concentric collection of large proliferating tumor cells located in a well oxygenated area in the periphery, while smaller non-proliferating cells populate deeper regions. The number of the "dormant" hypoxic tumor cells, which are far away from blood vessels than the normal diffusion distance of oxygen (about 100 μm to 150 μm), increase with spheroid size [19].

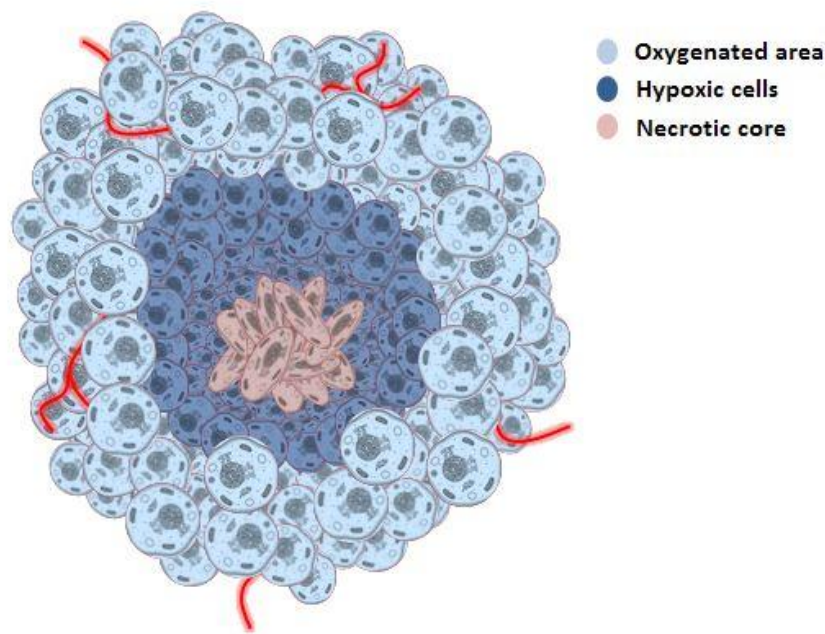

*Figure 1. Schematic representation of a multi-cellular solid tumor. Within the quasi-spherical shape, a concentric arrangement of aerobic and proliferating tumor cells are located in the periphery, while a smaller fraction of non-proliferating cells are concentrated in deeper regions, surrounding a core of dead cells (central necrotic core), emerging as a result of cell death due to diffusion limitations to oxygen and nutrients supply.*

As we already mentioned, the stroma and the neo-vasculature that arises provide solid malignant tumors with oxygen and nutrients necessary to survive and grow [8]. Nevertheless, the neo-vasculature blood vessels in the tumor are very different from normal vasculature. They are typically highly irregular and tortuous, have arterio-venous shunts and blind ends, lack smooth muscle or nerves and have incomplete endothelial linings and basement membranes [7]. As a result, the delivery of oxygen and nutrients to the tumor cells is much less efficient than in normal tissues and, therefore, the overall levels of oxygen in most tumors are low, resulting in areas with extremely hypoxic cells, surrounding the core area with anoxia. These hypoxic cells, developed in tumors because of the unbalanced growth of the tumor cells and of the vascular components needed to provide them with a sufficient blood supply, are radiation-resistant, limiting in some instance the effectiveness of the radiotherapy of cancer.

The hypoxic fractions of tumors are present in most of the human and experimental animal tumors, depending on tumor type, the degree of tumor differentiation, the tumor growth rate and the host tissue characteristics. Definitely, tumors contain a mixture of aerobic and hypoxic cells and this intra-tumoral heterogeneity is a common characteristic of solid tumors, depending on the cell type and on tumor microenvironment conditions. For more than sixty years ago, the tumor hypoxia impact to radiation therapy outcome is a headache for oncologists, as a possible cause for failure of local control of tumor response to radiation. Soon after the discovery of X-ray radiation and at the very beginning of experiments in radiation biology, it was reported by Schwarz in 1909 that the presence or absence of oxygen had a strong influence on the effectiveness of radiation to produce damage [20, cited in 21]. He demonstrated reduced skin toxicity when the skin was compressed (i.e. oxygen level reduced through expulsion of blood), although the specific role of oxygen was not known at that time. Moreover, the landmark papers of Gray *et al*. (1953) [22] and Thomlinson and Gray (1955) [23] pointed out the basic significance of the presence or absence of oxygen to the radiation response of malignant tissues undergoing radiotherapy. Thomlinson and Gray [23] postulated that, based on their histological studies on human lung adenocarcinomas, tumor cells would be forced away from blood vessels, beyond the effective diffusion distance of oxygen in respiring tissue, thereby becoming hypoxic and eventually necrotic. Furthermore, both experimental radiation biology research and clinical radiotherapy data, worldwide, show that, in the absence of oxygen, the dose of X-rays to produce the same degree of cell-killing as under well-oxygenated conditions would need to be increased by a factor of about 3 or even 5 in some tumors [24]. The increased killing of cells, observed in the presence of oxic conditions, is called "oxygen effect" and the quantitative expression of this phenomenon is given by the "oxygen enhancement ratio" (OER), defined as the ratio of doses under anoxic/hypoxic to aerobic conditions, to give the same cell survival or the same radiobiological effect. This is because oxygen molecules react rapidly with the free-radical damage produced by ionizing radiation in DNA, thereby 'fixing', or making permanent, the DNA damage that leads to cell death [22]. Several investigators support that hypoxic cells are also resistant to most anticancer drugs, as hypoxic cells are distant from blood vessels and, therefore, are not adequately exposed to some types of anticancer drugs.

Since hypoxia was recognized as an obstacle in cancer radiotherapy, several methods have been proposed to overcome it: a) hyperbaric oxygen therapy, b) neutron radiotherapy, c) hypoxic cells radiosensitization. The third approach is related to chemical modification of cellular radiosensitivity, by changing the mechanism of the overall radiolysis of the system. At this point, we have to mention another factor that critically influences the clinical outcome of cancer radiation treatment, specifically the intrinsic tumor radiosensitivity. Radiosensitivity is the probability of a cell, tissue or organ of suffering an effect per unit of radiation dose. There are several factors affecting the radiosensitivity, as for example physical factors (e.g. linear energy transfer, dose fractionation), chemical (e.g. oxygen, radio-protectors) and biological factors (e.g. cell cycle phase). According to the early observations of Bergonié and Tribondeau and their law stated at the beginning of the 20th century (1906), an increased sensitivity to radiation is observed in populations of cells that are highly proliferative (i.e. with rapid cell divisions) and less differentiated (i.e. stem cells are more sensitive than mature somatic cells). In other words, cellular radiosensitivity is proportional to the rate of cell proliferation and inversely proportional to the degree of cell differentiation. As a consequence, radiosensitivity of cells varies considerably as they pass through the cell cycle with, in general, late S-phase (synthesis of DNA) being most radio-resistant, very late G2 and mitosis most radio-sensitive and G1 phase taking an intermediate position [11, 25]. Substantial evidence through the decades of clinical cancer radiotherapy suggests that the intrinsic radiosensitivity of the tumor cells is the primary biological determinant of tumor response to radiation [26].

## 3. Current treatment strategies for cancer

Cancer is treated with both invasive and non-invasive treatment modalities, such as surgery, radiation therapy, chemotherapy, as well as other therapeutic modalities (e.g. immunotherapy, hormone therapy, biological therapy, photodynamic therapy) and a combination of these (e.g. radiosurgery). In oncology, the desired anti-cancer therapeutic clinical effect can be due to direct target tumor cell death by necrosis or apoptosis, vascular damage leading to tissue ischemia and resultant target cell death, or immune modulation, or a combination of these. Among the non-invasive treatment procedures, radiation therapy or radiotherapy is one of the most common treatments for cancer. Radiation therapy kills cancer cells that are dividing quickly and grow out of control, while it is desirable to avoid and/or minimize any damage to the surrounding normal cells. Traditional radiotherapy uses high energy electromagnetic waves, specifically photons (e.g. X-rays, gamma rays) or particles (e.g. electron beams, neutrons, carbon ions, alpha particles, and beta particles) to destroy tumor cells by damaging their DNA. Some ways that ionizing radiation can be given include: external beam radiation therapy, brachytherapy or internal radiation therapy, or use of injected radiopharmaceuticals. Radiation can be given alone or used in combination with other treatments, such as surgery or chemotherapy.

Chemotherapy uses chemical compounds to kill cells that divide rapidly, one of the main properties of cancer cells. This means that it also harms cells that divide rapidly under normal circumstances, as for example cells in the bone marrow, in the digestive tract and in hair follicles. In addition to pure chemotherapy, certain chemical agents (drugs), the so called radiosensitizers, can actually enhance cell response to ionizing radiation and they generally promote both the direct and indirect effects of radiation, facilitating the radiation to better destroy cancer cells, while side effects towards normal tissues are minimized. During the progress of the radiotherapy epoch, several chemical compounds (radiosensitizers) or physical agents (hyperbaric oxygen, hyperthermia) were developed to enhance the radiotherapeutic effect. Since the early pioneers in this field (G.E. Adams, J.F. Fowler, and J. Denekamp at Gray Cancer Institute in Mount Vernon Hospital, UK) [27], a plethora of research studies and applications were and are still performed on several radio-sensitization procedures to improve cancer radiotherapy [28]. Among them, hyperthermia is an adjuvant cancer treatment modality based on the cytotoxic properties of localized high temperatures (43 $^o$C) achieved within cancer cells and tissues. Hyperthermia is a treatment known from ancient times, but the renaissance of the method was mainly based on the modern electromagnetic heating techniques and is usually achieved by using microwaves to heat the tumor [14, 29].

Returning to radiotherapy of cancer, we have to spot that contemporary research and development in radiation oncology is mainly directed to further increase the treatment rates. This is currently achieved by the application of new ionizing radiation treatment modalities, such as intensity modulated radiotherapy (IMRT), intensity modulated arc-therapy, proton and heavy ion radiotherapy [11]. For instance, protons are already today used in radiotherapy to cure cancer by damaging irreversibly the DNA of cells, while allowing for a much-reduced collateral damage to surrounding healthy tissues [25], especially for deep-seated tumors. Proton therapy demonstrates this advantage because the physics of proton therapy is fundamentally different from the physics of conventional photons irradiation. Definitely, it is well documented that the basic difference between X-ray photons and particles is their different biological action and different depth-dose distribution. Photons show an exponentially decreasing energy deposition with increasing depth in tissue. In contrast, as pointed out first by Wilson in 1946, protons deposit their energy near the end of their path, leading to a maximum (the well-known "Bragg peak" phenomenon) near the end of range of the proton beam, without significant lateral scatter [30]. As a consequence, healthy tissues or adjacent critical organs located upstream and downstream of the tumor are practically unaffected by the irradiation. One can say that the curves of the distribution of the ionizing photons and the particles energy dose in depth behave as two mirrored curves [31]. Nevertheless, the enthusiasm for the medical applications of proton particles in oncology is somehow restricted by the uncertainties that can affect the accurate prediction of proton range *in vivo* [32] and by the high cost of proton treatment installations. Protons are accelerated conventionally by means of devices that use electric fields of two basic types: linear accelerators (linacs) and circular accelerators (e.g. cyclotron). Due to the above mentioned advantages of proton therapy, if ion accelerators were small and cheap as electron linacs, then probably radiotherapy would be made with protons, more precise and with the same therapeutic effects of γ-rays. However, because proton therapy is a costly treatment for cancer, today 99% of external beam radiotherapy is done with γ-rays and only a few centers in the world use "conventional" sources to perform proton-therapy, including approximately 8 centers in USA, 4 in Japan, 2 in China, 1 in Switzerland, 1 in Germany 1 in Korea, etc. According to "Particle Therapy Co-Operative Group", [http://ptcog.web.psi.ch/], for the year 2010, there were 31 active proton therapy centers, 2 heavy ion facilities, and approximately 12 neutron therapy centers in operation. There were as many as 22 new, planned, or proposed proton therapy centers, some of which today are already under construction. In some of these new centers it is proposed to combine both proton and heavy ion therapy facilities. Therefore, in nowadays all over the world, the advances in photon therapy continue to be competitive with the much more expensive hadrons therapy [33, 34]. Besides particles, high energy X-ray radiotherapy was considered as an alternative to cure deep-seated cancer. The treatment of pediatric tumors is one of the most important indications for proton therapy because a clear clinical benefit by means of reduced toxicity is assumed [25]. The high-energy photon beams, produced mostly by electron linacs, have energies of a few million electron volts, but are still called X-rays by medical doctors [35]. They have replaced low-energy X-rays and the gamma radiation from radioactive cobalt in some radiotherapy procedures because they deposit the dose (the energy per unit mass) at greater depth. For a qualitative representation of tissue simulator's depth dependence of the deposited dose, for different types of radiation, one can see the relevant figure, easily assessed on web (http://cds.cern.ch/record/1733872/files/vol46-issue10-p017fig.png?version=1), or in [35]. In the majority of the depth-dose distribution curves, liquid water target is usually chosen as a tissue simulator, due to the fact that biological tissues exhibit a large diversity and it is very difficult to correlate the published results. Nevertheless, significant deviations appear from water to real tissues, but water phantoms are still a common point of reference between radiotherapy centers worldwide. As an example, in Figure 2, the computer simulated depth-dose distribution (i.e. the Bragg peak position) is shown for a beam of protons, having initial incident energy of 200 MeV, in a target of liquid water, 1g/cm$^3$ density. The proton stopping power calculations were performed with the Monte Carlo computer code SRIM and the results show that protons of 200 MeV incident energy are characterized by an axial ion range of penetration, in liquid water, of Rm=256 μm. The statistical parameters of kurtosis and of asymmetry (skewness) that characterize this penetration depth are -0.8723247 and 51228 respectively.

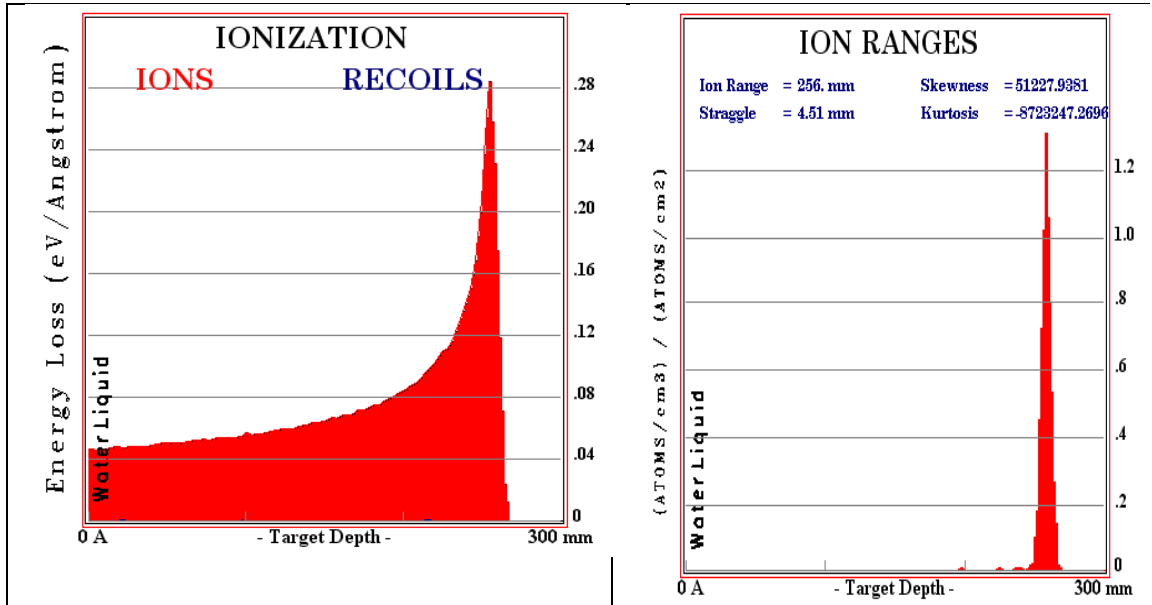

*Fig. 2. Depth dependence of the deposited energy dose in water, for protons of 200 MeV, with the narrow Bragg peak at the end.*

Without doubt, in anti-cancer fighting armamentarium, radiotherapy poses a principal position. On the other hand, radiotherapy based on ionizing radiation could not be the only option, as the high-energy photons releasing their energy into cancerous tumor can lead to significant damage to normal tissues surrounding the tumor. Therefore, advances in ionizing photon and particle radiation therapy modalities are still under investigation to enhance efficient and safe procedures. Simultaneously, there is a tendency to develop novel cancer treatment strategies based specifically on biophotonic principles, e.g. on the use of non-ionizing photons. In Biophotonics, the use of laser light photons, from ultraviolet to infrared part of the electromagnetic spectrum, gives totally new or older but revised biomedical applications in both diagnosis and therapy of human diseases. For example, powerful laser light beams are used instead of conventional scalpels for very careful and precise surgical work, including treating some cancers. Surgical excisions, performed with laser light, cause a non-contact, less bleeding (haemostasis) and damage to normal tissue than standard surgical tools do (mechanical or electrical scalpel), and there is a lower risk of infection and scarring. As another paradigm, photodynamic therapy (PDT) is a type of cancer treatment that uses visible or near infrared laser light to destroy tumors. Photodynamic therapy is a non-invasive or minimally invasive treatment technique, based on the simultaneous action of the following three factors, schematically represented in Figure 3: i) photosensitizing agents, able to photochemically eradicate malignant cells, ii) oxygen, generating highly reactive singlet oxygen, and iii) non-thermal monochromatic light irradiation. Briefly, in PDT, a certain drug, called photosensitizer or photosensitizing agent, is injected into a patient and distributed all over the patient's body. After a couple of days, the photosensitizer is preferentially accumulated in cancer cells. Laser light is then used to activate the agent and destroy cancer cells, in the presence of cellular oxygen [36].

In PDT, the light is usually monochromatic light, chosen to coincide with the maximum absorption wavelength of the photosensitizing agent molecules, in the red or near infrared part of the spectrum. The ideal light wavelength must give the proper photonic energy needed to photo-activate molecules for excitation and singlet oxygen production, as well as to ensure a good penetration of light in the tissue. These requirements impose the use of photons in the 600 – 850 nm part of the electromagnetic spectrum, since at wavelengths higher than 850–900 nm the photons might not have sufficient energy to produce $^1O_2$ and the most energetic photons in UV – VIS part of the spectrum have low penetration depth. The photo-generated singlet oxygen ($^1O_2$) attacks cellular targets, causing destruction through direct cellular damage, vascular shutdown, and activation of an immune response against targeted cells.

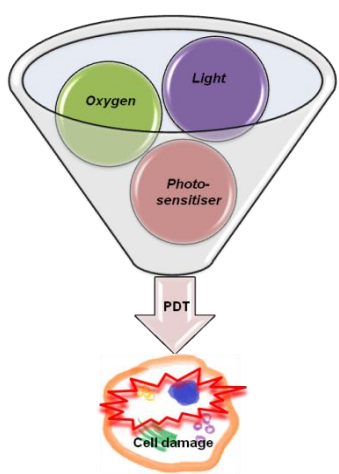

*Figure 3: In PDT, laser light activates a photosensitizing compound to destroy cancer cells, in the presence of cellular oxygen.*

We already mentioned that PDT is a non-invasive (especially in skin pathology) or minimally invasive (through flexible endoscopes in several internal organs) therapeutic procedure. Photodynamic therapy is also considered as a light-activated chemotherapy, which achieves a localized effect due to tumor selectivity of the photosensitizer, in contrast of traditional chemotherapy, which has a systemic effect. PDT is low cost, repeatable and less painful than conventional radiotherapy methods, while can also be effective for situations in which surgery is contra-indicated. Apart of PDT, laser light can be directly used to remove surgically cancer or precancerous tumors or to relieve symptoms of cancer, such as bleeding or obstruction.

The first demonstration of the photosensitization induced by ordinary light in oxygen-dependent chemical reactions and the term "photodynamic" were reported by Raab and von Tappeiner in 1904, over a century back. Nevertheless, the first attempts to use photosensitizing drugs for the cure of skin diseases dates back to ancient Egypt, India, and Greece, where psoralen-containing plant extracts and sun light (*heliotherapy*) were applied to treat psoriasis and *vitiligo*. On the other hand, the growing use of contemporary PDT is based on the pioneering work of Dougherty and associates [37], signalling the new PDT era. Indeed, considerable interest in phototherapy was regained since the discovery of the first laser by Maiman in 1960 and by the discovery of porphyrins as potential tumor localizers by several researchers. After the first laser-based clinical PDT report in late 1970's [37], the number of the relevant scientific research publications, for this rapidly expanding field, has been increasing exponentially. The history of PDT was presented in detail in several papers, as for example in a comprehensive review by Moan and Peng [38]. In the modern PDT era, this concept is still evolving rapidly and a great effort has been devoted towards the development of new photosensitizing agents, which exhibit specific light absorption and tissue distribution properties, as well as the development of cost effective monochromatic light sources, emitting in the photosensitizer absorption wavelength. In the last decades, PDT has begun to find approved uses in medicine and now is applied in oncology in many countries. However, photodynamic therapy has still some limitations in its clinical applicability, due to the necessity for accessibility to light of a deep located tumor.

Although PDT was originally developed as a tumor therapy, some of its most successful applications are also for non-malignant diseases [38]. Focusing in PDT clinical applications in oncology, in 1996 the FDA in USA approved PDT with hematoporphyrin derivative (HpD) as photosensitizer drug (Photofrin™) for palliative treatment of obstructive esophageal cancer, while two years later the FDA approved PDT with Photofrin™ for treatment of certain esophageal and obstructive lung cancers. A variety of other applications are evolving in clinical trials, including treatment of prostate cancer, head & neck cancer, skin cancer, pleural cancer, intraperitoneal cancer, and treatment of the resection bed after brain tumor surgical excision [39]. For the first twenty years of PDT, a review of this rapidly expanding field was presented in 1998 by Dougherty *et al* [36], receiving thousands of citations. Ten years later, in 2008, Wilson and Patterson published a comprehensive review paper [40], summarizing the status of PDT clinical applications with an emphasis on the contributions of physics, biophysics and technology, and the challenges remaining in the optimization and adoption of this treatment modality. Consequently, novel biophotonic modalities for both experimental and clinical PDT applications are in progress, on the subject of advanced light sources, new photosenitizers and optimization of both drug and light dose distribution in the lesion. Apart of the first generation of photosensitizers, nowadays there are new drugs that have been approved for PDT treatments. Porfimer sodium, temoporfin, 5-aminolevulinic acid and verteporfin are the most commonly used photosensitizers.

A very interesting problem, faced by several researchers and clinicians, is the complexity of PDT dosimetry due to the dynamic nature of the three, simultaneously acting, essential components—light, photosensitizer and oxygen, in order to control and optimize the photodynamic treatment efficacy. For example, an insufficient photosensitizer concentration in malignant tissue can lead to incomplete treatment of tumor, resulting in recurrence, while, in contrary, the photosensitizer overdose may cause significant damage to the healthy surroundings during the photodynamic treatment of the tumor [41]. Definitely, at least the first period of clinical PDT treatments was ruled worldwide by empirically prescribed dosimetry approaches. In an effort to develop a non-empirical monitoring method for PDT dosimetry, we measured in a preliminary work the absorption and laser induced fluorescence spectra in tissue phantoms, for identification and quantification of the concentration of the photosensitizer tetra(m-hydroxyphenyl)chlorin (m-THPC, Foscan$^{®}$) [42]. Regarding the dosimetry in photodynamic therapy, apart from the drug dose (e.g. the photosensitizer concentration in tumor), the light distribution in malignant tissue is also very important, which is governed by both the laser beam properties (e.g. wavelength, beam geometry, continuous wave or pulsed irradiation) and the optical parameters of the tissue (e.g. absorption coefficient, scattering coefficient, scattering anisotropy factor, and fluorescence quantum yield). The interdependence of these parameters and their implications for developing adequate dosimetry for PDT was examined in late 90's by Wilson *et al*. [43]. They termed as "explicit dosimetry" the traditional dosimetry approach, in which each dose factor (the administered photosensitizer dose, the incident light dose and the drug-irradiation time interval) is measured independently, *versus* to "implicit dosimetry", where the photosensitizer photobleaching is measured as an index of the effective delivered dose [43]. As we already mentioned, irrelevant to different photoactivated compounds used, many factors, such as oxygenation and vasculature, are influencing the cytotoxic outcome of PDT. Briefly, any oxygenation modification during PDT, resulting in oxygenation deficiency (i.e. hypoxia) or in photobleaching, as well as any chemical or physical factor that could modify the hemodynamics of the tissue may alter the photodynamic efficacy of PDT. More than a decade ago, we tried experimentally to follow the protective effect of ocreotide, a compound analog to somatostatin, supposing that ocreotide may have a modulating effect on tetra(m-hydroxyphenyl)chlorin (m-THPC, Foscan®) mediated photodynamic action in normal pig stomach, by decreasing the splachnic blood flow and/or suppressing the inflammatory process in the peptic system [41]. In modern PDT era, several methods are reported for the monitoring of the impact of photobleaching and hemodynamic responses to photodynamic therapy of cancer, targeting the effective dosimetry of PDT [44, 45].

Recently, in a mini review, we presented the outlines of the basic photophysical mechanisms that operate in PDT on a molecular and cellular level, as well as some of the new approaches in cancer therapy, concerning either the combination of PDT with radiotherapy, or the advances in drug delivery and the monitoring of the relevant photo-induced damage [46]. Definitely, many strategies have been proposed to maximize and potentiate the therapeutic effects of photodynamic therapy when used in combination with other curative modalities [47]. In this review, the authors report several applications of combined therapy in which PDT, being the core therapeutic partner, has been associated with both traditional and innovative therapeutic approaches for cancer treatment [47].

Regarding the cancer cells damage mechanism(s), it is noteworthy to spot the following paradoxical phenomenon: novel therapeutic strategies are developed to preferentially kill cancer cells through reactive oxygen species (ROS), although, in the same time, ROS are generally involved in cancer initiation (carcinogenesis), progression and metastasis. In nowadays, there are several research studies and therapeutical applications that answer this contradictory subject, addressed rhetorically by Wang and Yi as ''cancer cell killing *via* ROS – to increase or decrease, that is the question'' [48]. Therefore, the research efforts for optimization of cancer therapeutic strategies worldwide include the use of ROS-mediated cancer cells damage in photodynamic therapy, in combination with targeted drug delivery to secure selectivity. Moreover, as the major breakthrough in cancer management is the individualized patient treatment, new biophotonic techniques, e.g. photo-activated drug carriers, help the improvement of treatment efficacy and the obliteration of the normal tissue toxicity. For example, bioconjugated nanoparticles behave as Trojan horse for the delivery and targeting of anti-cancer drugs, taking advantage of increased ROS in cancer cells to enhance therapeutic activity and selectivity [46]. The use of non-systemic drug administration to the target, in conjunction with non-invasive or minimally invasive biophotonic therapeutic procedures (through flexible endoscopes), represents a very interesting personalized treatment option for patients. Indeed, any endoscope can be fitted with optical fibers or waveguides that transmit coherent (laser) or incoherent light and that light could precisely aim to cut surgically or destroy photo-dynamically a tumor.

In addition to PDT, recent studies support the possibility of photodynamic therapy and conventional radiotherapy combination. Concerning this combination, it is well known that the principal target for ionizing radiation-induced cell killing is nuclear DNA. Ionizing radiation causes DNA damage of cancerous cells by the generation of reactive oxygen species (ROS), especially DNA double strand breaks. Since PDT also induces several types of DNA lesions through singlet oxygen generation, there is a potential for synergism in cell killing due to interactions between the two modalities [49]. Moreover, radiosensitization of hypoxic tumor cells in combination with photosensitization of well oxygenated malignant tissue could eventually be a new alternative for optimization of cancer therapy [28]. This trimodality therapy, including drugs, ionizing radiation and visible light (in the electromagnetic spectrum wavelength range of 400 – 800 nm) results in additive or even synergistic effects on some lines of carcinoma cells [50]. However, although the double role of some chemical compounds as both a photo- and radio-sensitising agents promises unique perspectives for a concerted action of the two modalities to achieve a better control of tumor growth, the clinical outcome is not very well researched.

The last decades, with the rapid development of the nanotechnology, alternative methods have also been studied using nanosystems and nanomaterials for cancer treatment. Gold and ceramic nanoparticles, quantum dots, nanotubes, liposomes and dendrimers have been examined as drug delivery systems in personalized nanomedicine and the relevant pharmaceutical technology [46]. Ten years ago, Sinha *et al*. [50] reviewed the use of bioconjugated nanoparticles for the delivery and targeting of anticancer drugs. Then, they considered that the on-going efforts of scientists, researchers, and medical personnel and the future perspectives of nanotherapeutics in cancer, including the recent advances in proteomics and genomics, could sincerely ensure to ''*do big things using the very small*'' [50]. Without a doubt, nanotechnology, a flourishing scientific field that is everywhere and in everyday life, revolutionizes also cancer therapy by using nanoparticles as drug delivery systems for ROS generation.

Alongside to pure biophotonic procedures, briefly reviewed in [46], we anticipate the epitome of a multidisciplinary and transnational speculation in the last years, which suggests that lasers could also ameliorate the well established cancer proton therapy, offering a novel proton acceleration modality. With this step, the scientists exploit the use of photons from the entire electromagnetic radiation spectrum, from the ionizing part to the non-ionizing one, as a novelty for cancer therapy.

## **4. Bridging ionizing and non-ionizing radiation to cancer therapy**

As we already mentioned, proton therapy needs a high cost, very sophisticated technology requiring expensive facilities and, therefore, the relevant treatment units are still a few decades around the world. Almost 100,000 patients have been successfully treated with hadrons (protons and carbon ions) worldwide (PTCOG 2014, http://www.ptcog.ch/). Sophisticated proton therapy, based on conventional accelerators such as cyclotrons or synchrotrons, is now, and is likely to continue to be, more expensive than sophisticated (i.e., intensity-modulated) X-ray therapy, and the ratio of costs of proton vs. X-ray therapy per treatment fraction was estimated to be about 2.4 at 2005 [51]. State-of-the-art techniques borrowed from particle accelerators and detectors are increasingly being used in the medical field for the early diagnosis and treatment of tumors and other diseases. Novel technologies are under development, being expected to decrease the size and cost of proton therapy units and to increase the potential availability of proton therapy. Forty years after the discovery of laser light, the unique properties of the intense laser radiation brought new ideas in the acceleration of high-energy ion beams and a solid bridge between the ionizing and non-ionizing areas of electromagnetic radiation applications was established, as it is artistically illustrated in Figure 4. The interconnection of the possibilities of using an extended part of the electromagnetic spectrum in the biomedical applications field is definitely very attractive for novel, efficient and safer treatment strategies. At the beginning of 21$^{st}$ century, one century after the development of X-ray cancer radiation therapy, some laser-based future objectives for cancer radiotherapy were reported with great expectations. Their special properties make lasers versatile tools for biomedical applications that include very sophisticated anti-cancer therapeutical modalities. For example, as there was much expectation placed upon hadrons therapy, laser-driven proton acceleration was proposed to offer a potentially more compact (based on "table-top" laser systems) and significantly cheaper and convenient approach to generate a beam of protons for radiotherapy [52].

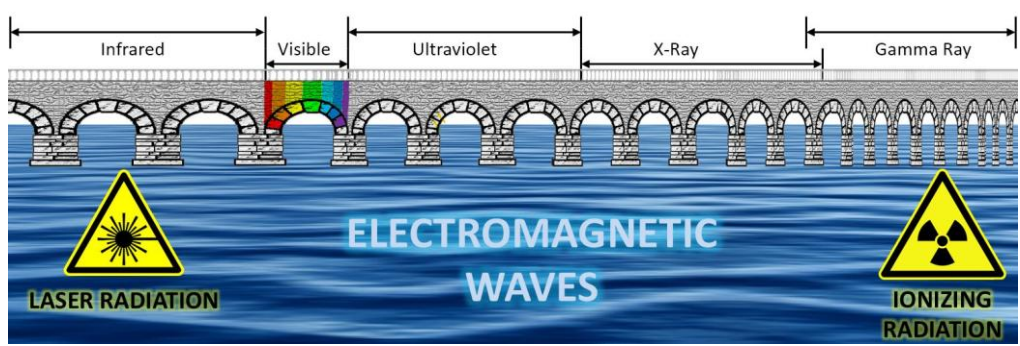

*Figure 4. Artistic visualization of a bridge between the ionizing and non-ionizing pillars of electromagnetic waves for cancer therapy.*

The conversion of high-energy radiation rays into visible light photons is a well studied and established field (i.e. photomultiplier tubes, scintillation detectors). On the other hand, the concept of conversion of optical photons to charged particles is a very innovative expectation, based on the unique properties of laser radiation and the distinct laser-matter interaction mechanisms. In several laboratories all over the world, the research in the area of laser-driven particles acceleration has grown dramatically since the first experimental evidence of multi-mega-electron-volt proton acceleration from laser-irradiated foils, and the number of related publications in the scientific literature grown-up exponentially [53]. Specifically, in 2005, a multinational team of researchers published a comprehensive review on the main experimental results and the relevant theoretical and computational models, regarding the fast ion generation by high-intensity laser irradiation of solid targets and their applications [54]. One year later, examples of the research efforts results from an ongoing research campaign on laser-driven proton acceleration at the Lund Laser Centre (LLC) in

Sweden were reported [54], highlighting the main mechanisms leading to ion acceleration in the interaction of high-intensity laser pulses with thin foil targets and the influence of the temporal contrast of the laser pulse on ions acceleration.

Taking a glance at laser-driven proton therapy (LPT) basic technology, we present briefly the procedure: Proton acceleration is achieved by focusing a high-power, pulsed laser on a thin foil target. The ultra short (sub-picosecond) laser pulses causes massive ionization and plasma creation in the target, expelling a large number of relativistic electrons. The sudden loss of electrons gives the target a high positive charge and this transient positive field accelerates protons to high energies [54]. A very common schematic representation of the LPT concept is illustrated in Figure 5.

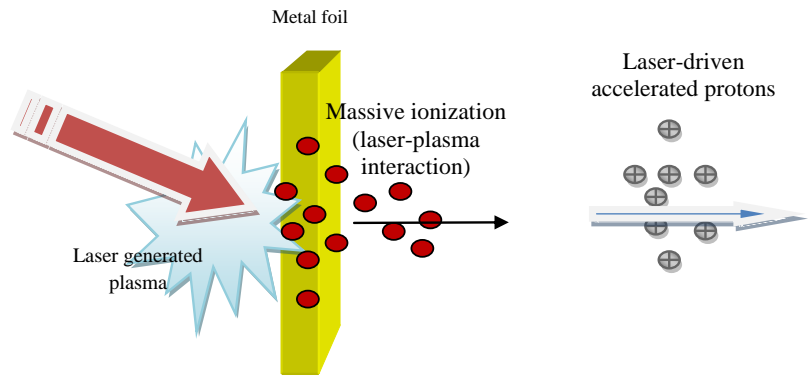

*Figure 5. Schematic representation of laser-driven proton acceleration for cancer therapy.*

As the researchers worldwide made a progress in the elucidation of the mechanism(s) for intense laser radiation and matter interaction, it was possible to produce high currents of protons with energies spanning the spectrum from few hundreds of electron volts up to several tens of MeV. Apart of the general idea for LPT, based on the laser induced massive ionization and plasma creation in the target, there are two particular mechanisms for proton acceleration with laser irradiation on a foil: Target Normal Sheath Acceleration (TNSA) and Radiation Pressure Acceleration (RPA) [55-58 and the literature cited in these papers]. In TNSA, protons are accelerated by the electric field created by the hot electrons heated by the laser. However, in most cases of TNSA with thick targets, the resulting ion energy spectrum is broad and only few protons reach the maximum energy, which is less suitable for applications requiring monoenergetic protons [59]. According to [59], it seems that RPA is a more efficient acceleration process for producing high energy quasi-monoenergetic protons, suitable for many applications, among them medical applications.

Nevertheless, the laser challenges for plasma physics and several interdisciplinary areas were addressed first to fast ignition research efforts, while further expansion of this field was envisaged to include also medical applications. The progression in laser matter interaction and a state of the art review regarding the high energy, solid-state Petawatt lasers (Petawatt, 1 PW = $10^{15}$ W) for fast ignition research was reported in 2006 [60]. When laser pulses of such immense power are focused onto beam pot as small as a few $\mu m^2$ only, light intensities in the order of $10^{20}$ or $10^{21}$ W/cm$^2$ can be achieved even in a university laboratory. During the first 21$^{st}$ century decade, a number of new laser systems were under construction or were planned to develop around the world, which would facilitate the laser-driven hadrons acceleration experiments. From the already developed high-energy Petawatt lasers in large laser facilities, we expect that the new generation of high power lasers will increase the accelerated hadrons energy by orders of magnitude, culminating in the pan-European Extreme Light Infrastructure (ELI) project for ultrafast laser beamlines facilities, in which an Exawatt laser is expected to reach $10^{26}$ W/cm$^2$ in a few years [61]. At intensities of this order charged particles traversing a laser wavelength gain huge electromagnetic energies such that even protons become relativistic. The goal of the new "Extreme Light Infrastructure –ELI" European facilities is to yield the highest peak power and laser focused intensity and the expectation for the regime beyond the exawatt, towards the zettawatt is the next high intensity frontier [62].

The acceleration of ions by laser irradiation of foils has been pursued actively using experiments, theory and numerical simulations. The experimental implementation of laser-driven particles acceleration has been performed by several research teams, in multi-national laser facilities, under very specific physical conditions and using pulsed laser systems with different characteristics (e.g. wavelength, ultrafast pulses, and light intensity).

The majority of the recently reported studies support that lasers could ameliorate cancer proton therapy [61, 62]. It is reported that laser-driven proton acceleration offers a significantly cheaper and relatively convenient approach to generate a beam of protons and radiobiological experiments in this direction indicate that protons in laser proton therapy (LPT) kill cancer cells as effectively as those from accelerators. Recently, monolayers of human cervical cancer cells (HeLa) were exposed to laser-driven quasi-monoenergetic protons, generated with a table-top laser system, delivering single shot doses up to 7Gy to living cells [63]. Although the biophysical studies for laser-driven ion acceleration are still in their infancy, very promising radiobiological processes are reported, since protons and photons have a parallel destructive effect on the DNA of the targeted tumor cells [63].

Recently, the global progress toward laser-driven radiotherapy is comprehensively reviewed in [61]. In this paper, the authors address the question if and how the radiobiological effectiveness (RBE) generated by laser-driven particle acceleration differs from that provided by conventional accelerators [61]. They present briefly some protons therapy centers, in operation and planned, based mainly in multidisciplinary transnational collaboration efforts that explore the development of laser-driven acceleration technology and the potential biomedical applications. Nevertheless, they conclude that before answering on the effectiveness of laser-driven hadrons therapy, we have to wait at least six years for completion of the first randomized trial that is testing the efficacy of conventional accelerator driven proton therapy and also an estimated time frames of more than 3–10 years for implementing laser driven ion beam radiotherapy [61].

## **Concluding remarks**

In nowadays, the anti-cancer edifice has four main pillars: surgery, radiotherapy, chemotherapy and immunotherapy. Surgery and radiotherapy are considered as localized forms of treatment, while chemotherapy and immunotherapy are systemic treatment forms that may be used alone or in combination with other therapeutic modalities. As the long-standing dream in the field of cancer therapy is the ability to precisely treat deep-seated tumors noninvasively, it is imperative to find new diagnosis and therapy platforms driving this idea.

In this review, we came across at the challenges that a broad band of electromagnetic radiation offers to cancer fighting methodologies. As we mentioned, ionizing radiation therapy of cancer is mainly considered as a local treatment modality, since the radiation energy is targeting well-defined areas of the body, where there is the tumor volume. Nevertheless, there is a risk of unwanted irradiation of healthy tissue surrounding a targeted tumor that may cause secondary carcinogenesis or may promote invasion and migration of cancer cells, leading to metastasis. Moreover, not all the forms of radiation therapy can reach safely and effectively all parts of the body, especially if cancerous cells have spread to distant areas. Therefore, a new approach causing selectivity to malignant cells and no damage to tissue surrounding tumors is needed. At this point, the revolution in the fields of monochromatic and coherent laser light physics and technology gave to the oncology stage new diagnosis, therapy and monitoring modalities. The use of laser scalpel for cancer surgery and new biophotonics treatment techniques, such as photodynamic therapy and photo-activated drug carriers were introduced in oncology to improve cancer treatment efficacy and concomitantly to reduce normal tissue toxicity. Definitely, PDT is an emerging cancer-treatment modality. Moreover, the possibility that laser could not only fire directly at the malignant tissue, but can also be used to irradiate thin foil targets for laser-driven particle acceleration gives notable perspectives in cancer therapy. According to Ledingham *et al.* [61], the overall challenge for laser-driven ion beam radiotherapy is "to develop well-controlled, reliable energetic ion beams of very high quality that can meet stringent medical requirements with respect to physical parameters and performance", representing a viable alternative in the 21$^{th}$ century' radiotherapy.

Facing cancer in one site and the impact of the novelties in the electromagnetic radiation therapy field in the other site, we must take in consideration that the majority of the involved research teams are primarily trained in physics, chemistry, mathematics, pharmacology, nanotechnology and engineering and that, eventually, they require an interfacing bridge between technology and clinics to integrate their scientific advances in clinical use. On the other hand, clinicians classically are not predisposed to attempt into advanced technology research, suspending the implementation of sophisticated technology improvements for a "bench-to-bedside" approach. The history of clinical and experimental cancer radiotherapy reflects this interconnection of basic science and meaningful health outcomes in medical practice. Moreover, the latest progress of nanotechnology signed new perspectives in oncology and new diagnostic and therapy platforms, driving the expansion of personalized medicine.

The basic idea of personalized or targeted cancer therapy is to deliver the precise dose of a drug to the individual patient, at the right time interval, according to his tissue-response endpoints. Therefore, the scope of personalized medicine is broad and aims to encompass the complex biology intrinsic to most cancers. Hopefully, the possibility of data handling and processing, by using high performance computing and informatics, ensures the draw-on of novel cancer monitoring treatment modalities, despite the plethora and the heterogeneity of data on cancer biology processes. In this aspect, in combination with the emergent scientific, technological and medical discipline of *in silico* oncology, computational cancer models are expected to enhance individualized treatment optimization, through experimentation *in silico*, i.e. on the computer [16]. Without doubt, as a major breakthrough in cancer management is the individualized patient treatment, the new biophotonic techniques, e.g. photo-activated drug carriers, could definitely help the improvement of treatment efficacy and/or the elimination of normal tissue toxicity. For example, considering the experimental achievements of optical trapping and manipulation of cells, we can extend the possibilities of firing certain regions within a cell, which is successfully manipulated by optical forces, using a sophisticated laser micro-scalpel method, to irradiate a specific cell or region of a cell, instead of the entire cell population of a tumor.

In conclusion, trying to predict what does the future hold for cancer therapy, it seems that the gold era of cancer radiation therapy in 21$^{st}$ century will be marked by the exploitation of the entire spectrum of electromagnetic waves, in order to expand the interface between imaging and therapy and to devise effective treatment strategies for all types and stages of cancer, in benefit of humanity.

**Acknowledgments**
The author expresses heartfelt thanks to all her colleagues and collaborators, both past (during her Ph.D. thesis in Carol Davila Medical University of Bucharest, Romania) and present, for their contributions that led to fascinating passes through the electromagnetic spectrum. Special thanks to Prof. M. Kokkoris and his group for the proton stopping power calculations and to her Ph.D. student A. Gousetis for part of the graphical work.